\documentclass[10pt,twocolumn,twoside,letterpaper]{IEEEtran}
\pdfoutput=1 

\usepackage{geometry}
\makeatletter
\def\ps@IEEEtitlepagestyle{
  \def\@oddfoot{\mycopyrightnotice}
  \def\@evenfoot{}
}
\def\mycopyrightnotice{
  {\footnotesize
  \begin{minipage}{\textwidth}
  \centering
  978-1-7281-4164-0/19/\$31.00 \copyright2019 IEEE
  \end{minipage}
  }
}

\usepackage{url}
\usepackage{xcolor}
\usepackage{upgreek}
\usepackage{amsmath}
\usepackage{amssymb}
\usepackage{graphicx}
\usepackage{hyperref}

\begin{document}

\title{Large-area Si(Li) Detectors for X-ray Spectrometry and Particle Tracking for the GAPS Experiment
}
%
%
%

\author{Field Rogers, 
	   Mengjiao Xiao, 
	   Kerstin Perez, 
	   Steven Boggs, 
	   Tyler Erjavec, 
	   Lorenzo Fabris, 
	   Hideyuki Fuke, 
	   Charles J.~Hailey, 
	   Masayoshi Kozai, 
	   Alex Lowell, 
	   Norman Madden, 
	   Massimo Manghisoni, 
	   Steve McBride, 
	   Valerio Re, 
	   Elisa Riceputi, 
	   Nathan Saffold, 
	   Yuki Shimizu, 
	   and Gianluigi Zampa
\thanks{Manuscript received December 12, 2019.  }
\thanks{F.R., M.X., K.P., and T.E.\ are at the Department of Physics, Massachusetts Institute of Technology, Cambridge, MA 02139 USA (email: frrogers@mit.edu).}
\thanks{S.B.\ is at the Department of Physics, and A.L\ is at the Center for Astrophysics and Space Sciences, at the University of California San Diego, La Jolla, CA 92093 USA.}
\thanks{L.F.\ is at Oak Ridge National Laboratory, Oak Ridge, TN 37831 USA.}
\thanks{H.F.\ and M.K. are at the Institute of Space and Astronautical Science (ISAS) at the Japan Aerospace Exploration Agency (JAXA), Sagamihara, Kanagawa 252-5210, Japan.}
\thanks{C.J.H., N.M., and N.S.\ are at Columbia Astrophysics Laboratory, Columbia University, New York, NY 10027, USA.}
\thanks{M.M., V.R., and E.R.\ are jointly at the Department of Engineering and Applied Sciences, University of Bergamo, Dalmine I-24044, BG, Italy and INFN, Sezione di Pavia, Pavia I-27100, Italy.}
\thanks{S.M.\ is at the Space Sciences Laboratory, University of California at Berkeley, Berkeley, CA 94720, USA.}
\thanks{Y.S.\ is at the Department of Physics, Faculty of Technology, Kanagawa University, Yokohama, Kanagawa 221-8686, Japan.}
\thanks{G.Z.\ is at INFN presso l'AREA di Ricerca di Trieste, Trieste I-34149, Italy.}
}

\maketitle

\pagenumbering{gobble}

\begin{abstract}
Large-area lithium-drifted silicon (Si(Li)) detectors, operable 150$^{\circ}$C above liquid nitrogen temperature, have been developed for the General Antiparticle Spectrometer (GAPS) balloon mission and will form the first such system to operate in space. These 10\,cm-diameter, 2.5\,mm-thick multi-strip detectors have been verified in the lab to provide $<$\,4\,keV FWHM energy resolution for X-rays as well as tracking capability for charged particles, while operating in conditions ($\sim$-40C and $\sim$1 Pa) achievable on a long-duration balloon mission with a large detector payload. These characteristics enable the GAPS silicon tracker system to identify cosmic antinuclei via a novel technique based on exotic atom formation, de-excitation, and annihilation. Production and large-scale calibration of $\sim$1000 detectors has begun for the first GAPS flight, scheduled for late 2021. The detectors developed for GAPS may also have other applications, for example in heavy nuclei identification.
\end{abstract}


\section{Introduction}

The first lithium-drifted silicon (Si(Li))~\cite{Spieler, Semikon} detectors to satisfy the unique geometric, performance, and production requirements of the General Antiparticle Spectrometer (GAPS) experiment have been produced and their performance validated and understood. GAPS is designed to detect low-energy ($<$0.25 GeV/n) cosmic antinuclei, in particular rare antideuterons that could be signatures of dark matter, using a novel exotic atom-based particle identification scheme on three Antarctic balloon missions ($\sim$3 months total exposure)~\cite{AntideuteronAramakiSensitivity, AntiprotonAramakiSensitivity}. In this contribution, we present the X-ray energy resolution and particle track reconstruction capabilities of the GAPS Si(Li) detectors.

\section{Requirements and Specifications}
The GAPS instrument consists of a plastic scintillator time-of-flight (TOF) surrounding a tracker composed of $\sim$1000 Si(Li) detectors arranged in 10 planes. 
A cosmic antinucleus first traverses the TOF, which measures velocity and energy deposition and is the basis of the system trigger. The particle then loses energy as it moves through the Si(Li) tracker, where it is eventually captured by a silicon nucleus, forming an exotic atom in an excited state. Particle identification is based on the de-excitation X-rays and hadronic annihilation products of the exotic atom in the tracker, as well as d$E$/d$x$ and stopping depth information. 
The Si(Li) system is critical to the success of the GAPS experiment as it serves as the target for the incoming low-energy antinucleus, measuring its energy deposition and stopping depth in the instrument, and capturing the antinucleus into an exotic atom; the spectrometer for X-rays from the de-excitation of the exotic atom; and the tracker for the products of the nuclear annihilation. The detectors must provide the large geometric acceptance needed for a rare-event search and operate with a low-power readout. Since the large instrument size precludes the use of a cryostat or pressure vessel, the detectors are designed to operate at flight pressures and at temperatures $\sim$-40$^{\circ}$C, achievable using an oscillating heat pipe system~\cite{OHP} for cooling. Since the GAPS design calls for $>$\,1000 detectors, a low-cost fabrication method with high yield is also required. 

\begin{figure}[htbp]
\centerline{\includegraphics[width=.42\textwidth,trim=100 20 130 295,clip]{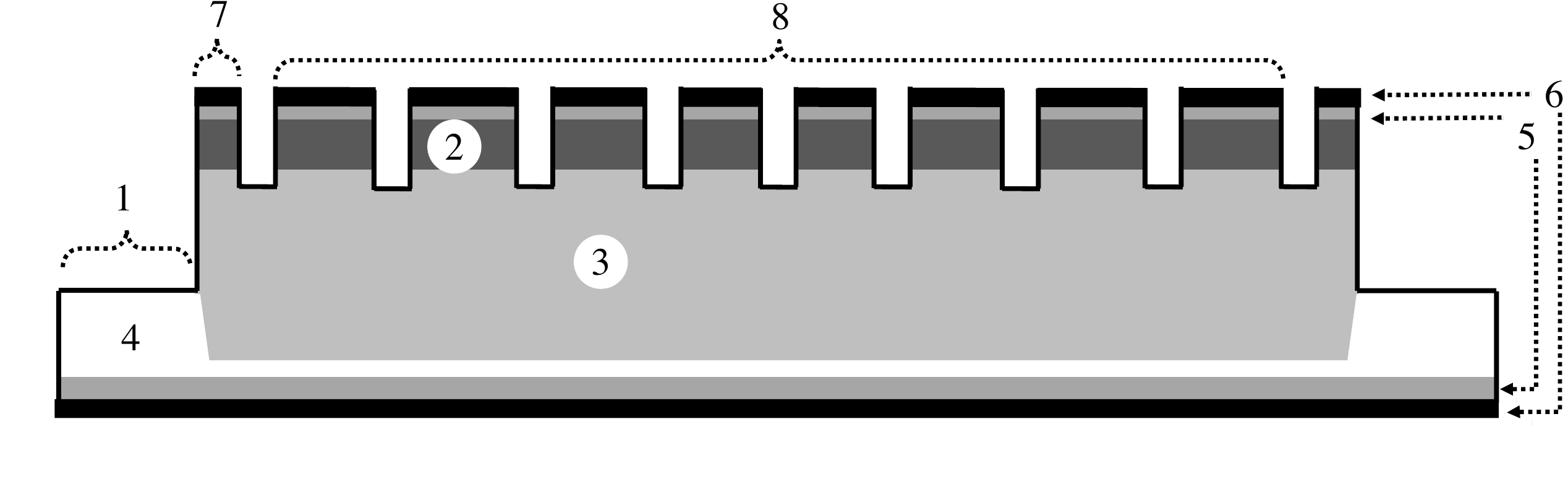}}
\centerline{\includegraphics[width=.42\textwidth,trim=870 900 1290 270,clip]{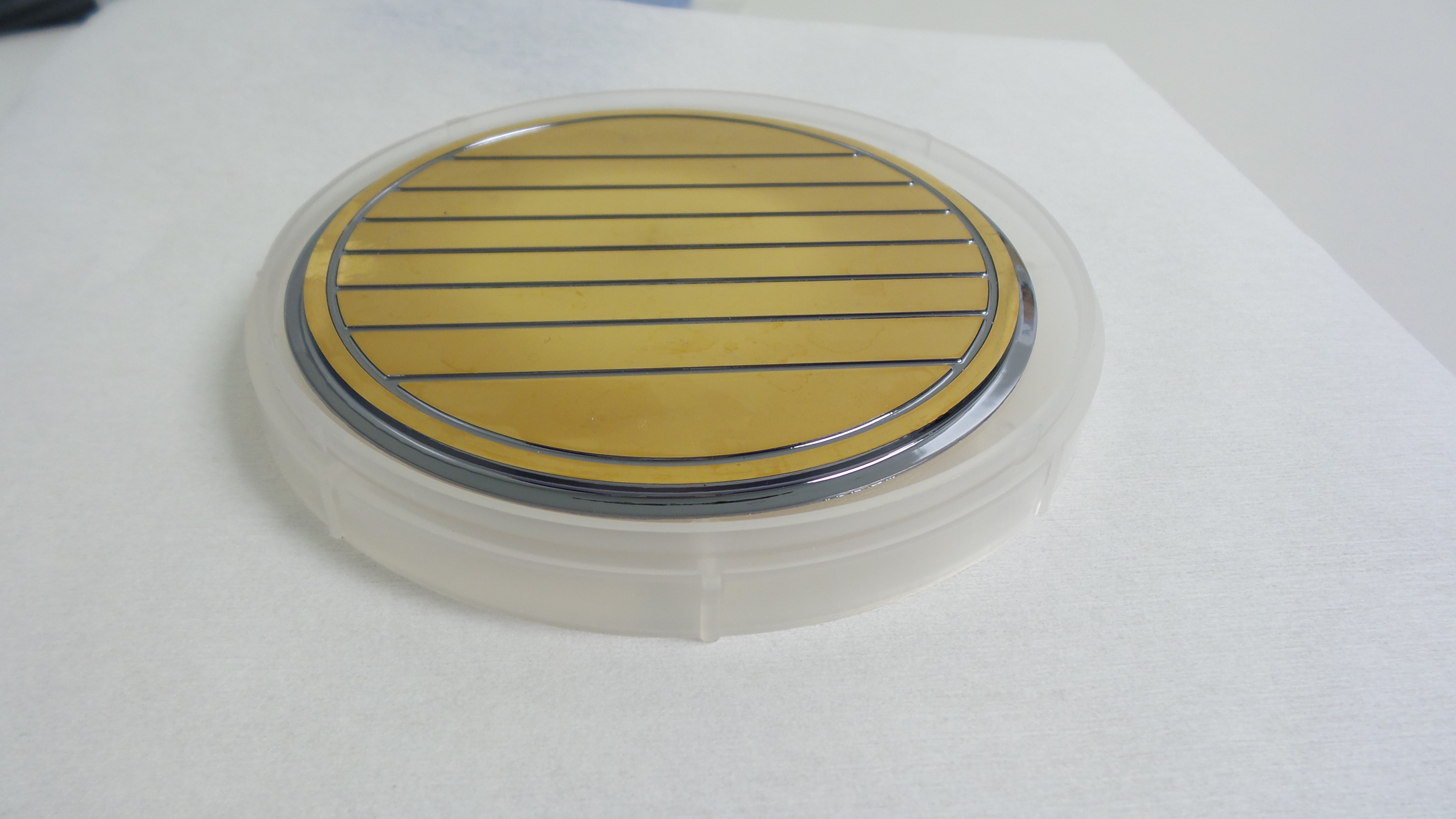}}
\caption{
\emph{Top}: 
Cross-sectional diagram of a GAPS detector (not to scale). Drifting is facilitated by a $\sim$1~mm-deep, $\sim$3~mm-wide top-hat brim (1). Li ions from the $\sim$0.1~mm-thick \emph{n}$^+$ Li-diffused layer (2) are drifted through the \emph{p}-type wafer to form a compensated active volume (3). The top hat brim and a 0.1~mm-thick \emph{p}-side remain undrifted (4). The electrical contacts consist of $\sim$20~nm Ni (5) and $\sim$100~nm Au (6). The $\sim$1~mm-wide, $\sim$0.3~mm-deep grooves separate the guard ring (7) from the active region (8) and segment the active region into parallel strips of equal area and capacitance. 
\emph{Bottom}: Photograph of a GAPS flight detector.
}
\label{fig:det} 
\end{figure}

The GAPS Si(Li) fabrication method, which has a yield $>90$\%, was developed in collaboration with Shimadzu Corporation~\cite{PerezSiLi, KozaiIEEE, KozaiNIM} and is based on a silicon substrate developed in collaboration with SUMCO Corporation. The materials cost per detector is $\sim$\$500. The geometry for the GAPS flight detectors is shown in Fig.~\ref{fig:det}. The 2.5\,mm depth is thick enough to stop a 0.25\,GeV/n antinucleus in $<$10 layers, but thin enough to provide high escape fractions for $20-100$~keV de-excitation X-rays. With a diameter of 10 cm, $\sim$1000 devices are needed to fill the ten 1.6\,m $\times $1.6\,m planes of the tracker. 

An energy resolution of $\lesssim$4~keV FWHM in the $20-100$~keV range is required for discrimination between de-excitation X-rays of different exotic atom species and is thus fundamental to the GAPS particle identification scheme. At the same time, the readout electronics must operate on a limited power budget to be compatible with GAPS' balloon nature. In light of the limited power for readout electronics, the Si(Li) system can reach the required energy resolution by controlling detector leakage current and capacitance, two of the dominant detector-related noise sources as discussed in section~\ref{xrays}. The low leakage current is achieved at relative high temperatures ($\sim-40^{\circ}$C) by implementing a thin undrifted layer on the detector's \emph{p}-side and a guard ring structure (Fig.~\ref{fig:det}) that limits surface leakage current in the detector's active area~\cite{KozaiNIM}. The capacitance and leakage current are further controlled by segmenting the detectors into strips, which also improves spatial resolution for particle tracking. With 8 strips per detector, the per-strip capacitance of $\sim$40~pF is sufficiently low that the required energy resolution can be achieved with a low-power application-specific integrated circuit (ASIC). The ASIC power of $\lesssim10$\,mW per channel is low enough to read out 8 channels for each of $>1000$ Si(Li) detectors while staying within the overall 110\,W power budget for readout electronics, and a custom ASIC will be used for Si(Li) detector readout during flight~\cite{Manghisoni, Scotti}.

Detectors that meet these requirements may also be suitable for other applications, e.g., identification of heavy nuclei at rare isotope facilities such as the National Superconducting Cyclotron Lab or Facility for Rare Isotope Beams~\cite{NSCL_Beta,NSCL_frag}. A few GAPS detectors would provide sufficient stopping
power for relevant nuclei, and fine spatial resolution is
not required.

\section{Si(Li) Detector Performance}
\subsection{Laboratory Testing Setup}
\label{sec:setup}
Details of detector preparation and experimental methods are 
in~\cite{Rogers_JINST}. In short, the detector's \emph{p}-side is biased at $-250$~V relative to the guard ring and signal processing electronics. While the ASIC is under development, each strip is read out from the \emph{n$^+$}-side by a discrete-component charge-sensitive preamplifier~\cite{LorenzoPreamp}. Temperature is monitored with a calibrated diode. 

Energy resolution in the 20--100\,keV range is assessed in an aluminum vacuum chamber at $\sim$1~Pa, with the detector cooled by flowing nitrogen in a closed system. Two $\gamma$ lines are used, 59.5\,keV ($^{241}$Am) and 88.0~keV ($^{109}$Cd). Signal from a preamplifier channel is shaped by a Canberra 2020 Spectroscopy Amplifier with variable peaking time and digitized by an Ortec Ametek Easy MCA module. 

The cosmic muon measurement is recorded in a nitrogen atmosphere cooled by injecting cold nitrogen. To eliminate non-muon background, coincident hits are required between the corresponding strips of two vertically-stacked detectors. The preamplifier signals are processed by a CAEN N6725 digitizer, enabling coincident triggering. A rough calibration is extrapolated from the 59.5~keV peak of $^{241}$Am. 

\subsection{Tracking Performance}
The response of a strip at $-40^{\circ}$C to heavy particle tracks is assessed using cosmic muons. 
The data in the top panel of Fig.~\ref{fig:performance} show the cosmic muon spectrum obtained by the method in section~\ref{sec:setup}, which is consistent, given the uncertainty in calibration, with the expected distribution of energy deposition in 2.3~mm of active silicon depth. 

Antinuclei in the GAPS energy range are too slow to be
MIPs and therefore will deposit more energy as they traverse
the detectors. Different energy deposition signatures can be
used for identification of incident particles as
they slow to stop from up to 0.25 GeV/n. Accordingly, the ASIC is
designed to accept signals up to 100 MeV before saturating.

\subsection{Energy Resolution Performance}
\label{xrays}
\begin{figure}[htbp]
\centerline{\includegraphics[width=.41\textwidth,trim=32 0 7 0,clip]{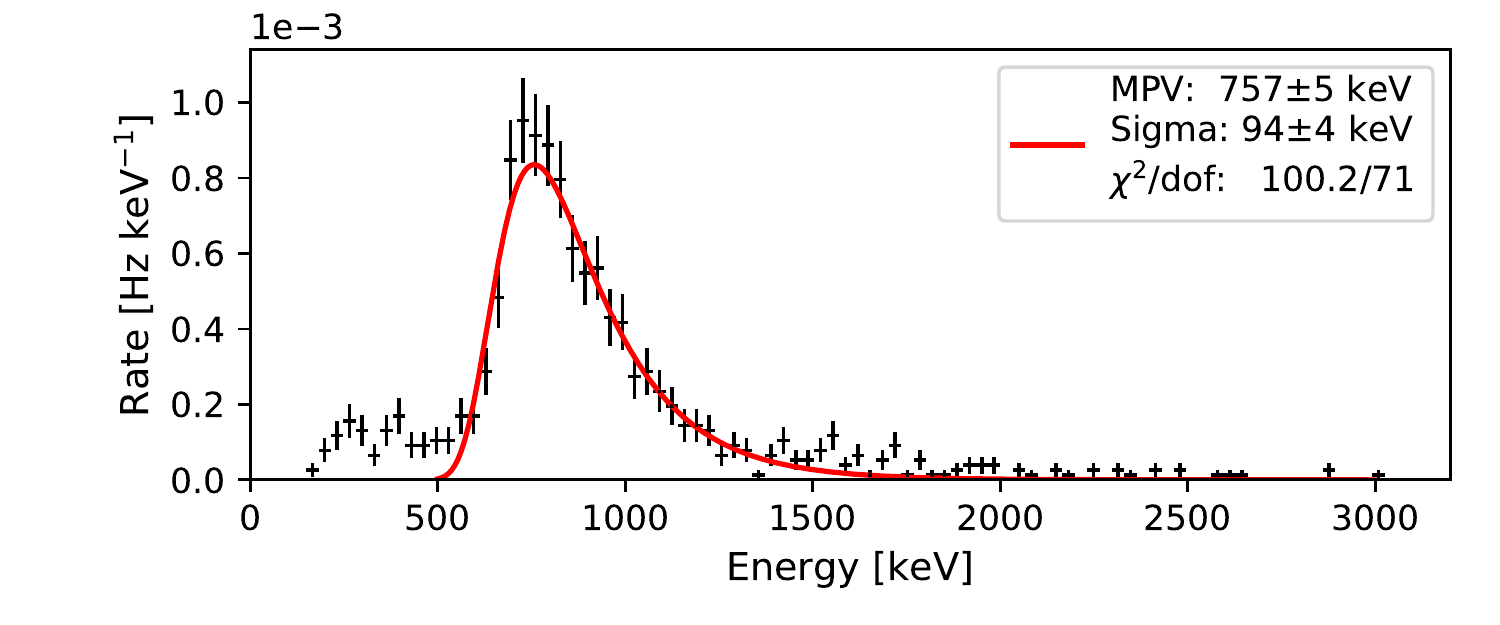}}
\centerline{\includegraphics[width=.41\textwidth,trim=16 0 7 8,clip]{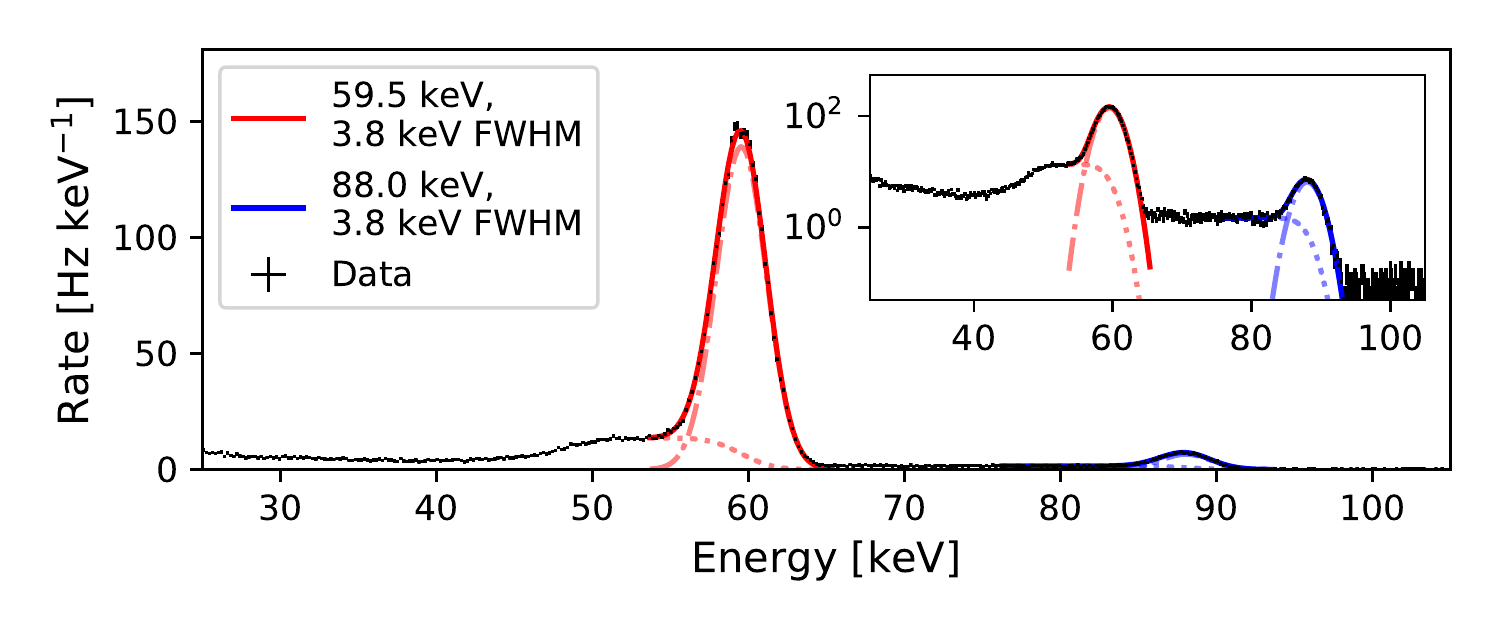}}
\centerline{\includegraphics[width=.41\textwidth,trim=1 10 7 8,clip]{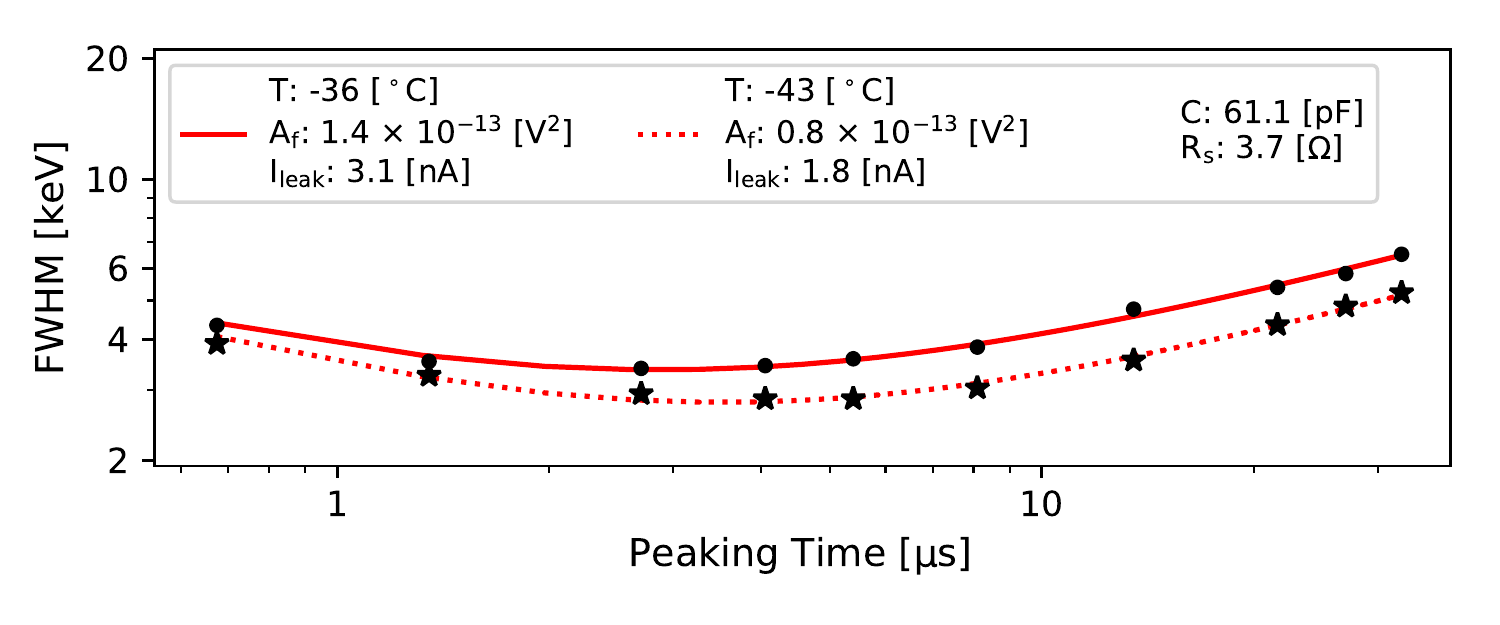}}
\caption{\emph{Top}: Spectrum of cosmic muons. A Landau distribution (red) fitted to the data describes the fluctuations of energy loss in the detector. \emph{Middle}: Spectrum of $^{241}$Am and $^{109}$Cd around their 59.5 and 88.0~keV peaks, with fits to the photopeak and Compton scattered components. The inset shows the same data in semi-log format to better display the 88.0~keV peak. \emph{Bottom}: Energy resolution at 59.5~keV varies with peaking time at two temperatures. The best fit of \eqref{eq:model} is shown in red.}
\label{fig:performance} 
\end{figure}

The response of a GAPS detector strip irradiated by $^{241}$Am and $^{109}$Cd is shown in the middle panel of Fig.~\ref{fig:performance}. Each mono-energetic line manifests
as a Gaussian photopeak and a nearly-flat low-energy tail
from Compton scattering from surrounding materials. $<$\,4~keV FWHM was achieved at the relatively high temperature of $-35^{\circ}$C.

To understand the different noise contributions to the energy resolution, we use the established noise model~\cite{Goulding, Spieler} for a semiconductor read out by a discrete-component preamplifier and shaping electronics: 
\begin{equation}
ENC^2\!\! =\!\! \Big(2qI\!\!+\!\!\frac{4kT}{R_p}\Big)\tau F_i\!+\!4kT\Big(R_s\!\!+\!\!\frac{\Gamma}{g_m}\Big)\frac{C^2}{\tau}F_{\nu}\!+\!A_fC^2F_{\nu f}, \nonumber
\end{equation}
\begin{equation}
FWHM\!=\!2.35(\epsilon/q)\times ENC.
\label{eq:model}
\end{equation}

In \eqref{eq:model}, $q$ is the electron charge, $k$ is Boltzmann's constant, $T$ is temperature, and $\epsilon$ is the ionization energy of silicon. $I$ and $C$ are the leakage current of the strip and total capacitance, the latter having strip, preamplifier, and stray components. The parallel resistance $R_p$, scaling factor $\Gamma$, and transconductance $g_m$ are preamplifier properties while multiple sources may contribute to the series resistance $R_s$ and coefficient of $\frac{1}{f}$ noise $A_f$. $\tau$ is the peaking time of the shaper and $F_i$, $F_{\nu}$ and $F_{\nu f}$ are shaper-dependent coefficients of the different noise terms. 
Detector contributions to $I$ and $C$ are measured independently. 

In the bottom panel of Fig.~\ref{fig:performance}, the energy resolution of the 59.5~keV line is recorded at two temperatures and a range of peaking times for a single detector strip, and \eqref{eq:model} is fit to the data using the fitting method detailed in~\cite{Rogers_JINST}. The energy resolution as a function of $\tau$ and $T$ is well-described by the noise model, lending us confidence in predicting energy resolution with different readout electronics or at varying flight temperatures.

\section{Conclusions}
Si(Li) detectors meeting the unique requirements of the GAPS experiment have been produced by Shimadzu Corporation and their tracking and energy resolution performance have been tested and understood in the laboratory setting. In
particular, the detectors exceed the requirement of $<$\,4~keV
FWHM energy resolution at $-40^{\circ}$C, when read out with custom
discrete preamplifier electronics in the laboratory setting. In flight,
they will  be coupled with a custom ASIC for readout. Large-scale production and calibration are underway: since January 2019, the GAPS collaboration has received $\sim$70 detectors monthly, and $\sim$1000 detectors will operate on the first GAPS flight scheduled for December 2021.

\section*{Acknowledgment}
We thank SUMCO Corporation and Shimadzu Corporation for their cooperation in detector development. 
We also thank our GAPS collaborators for their consultation and support.

K.\ Perez receives support from the Heising-Simons Foundation and the Alfred P.\ Sloan Foundation. F.\ Rogers is supported through the National Science Foundation Graduate Research Fellowship under Grant No.\ 1122374. M.\ Kozai is supported by the JSPS KAKENHI under Grant No.\ JP17K14313. H.\ Fuke is supported by the JSPS KAKENHI under Grant Nos.\ JP2670715 and JP17H01136. M.\ Manghisoni, V.\ Re, E.\ Riceputi, and G. Zampa are supported by the Italian Space Agency through the ASI INFN agreement n. 2018-28-HH.0: "Partecipazione italiana al GAPS - General AntiParticle Spectrometer".

This work was supported in part by the NASA APRA program through Grant Nos. NNX17AB44G
and NNX17AB46G.

\bibliographystyle{ieeetr}

\end{document}